\documentclass[aps,prc,twocolumn,amsmath,superscriptaddress,nofootinbib]{revtex4-1}

\usepackage{varwidth}
\usepackage{multirow}
\usepackage{dcolumn}
\usepackage{tabularx}
\usepackage{booktabs}
\usepackage{makecell}
\newcolumntype{C}{>{\centering\arraybackslash}X}

\usepackage{graphicx}
\usepackage{epstopdf}
\usepackage{graphics}
\usepackage{graphicx}
\usepackage{epsfig}
\usepackage{epstopdf}
\usepackage{hyperref}
\hypersetup{
    colorlinks=true,
    linkcolor=blue,
    citecolor=blue,
    urlcolor=blue,
    anchorcolor=blue}

\begin{document}
\title{A novel approach for studying two-particle momentum correlation
function in relativistic nuclear collisions}

\author{Zhi-Lei She}
\email[]{shezhilei@wtu.edu.cn}
\affiliation{School of Mathematics and Statistics, Wuhan Textile University, Wuhan 430200, China}

\author{Wen-Chao Zhang}
\affiliation{School of Physics and Information Technology, Shaanxi Normal University, Xi'an 710119, China}

\author{An-Ke Lei}
\email{ankelei@mails.ccnu.edu.cn}
\affiliation{School of Physics and Electronic Science, Guizhou Normal University, Guiyang 550025, China}

\author{Dai-Mei Zhou}
\email[]{zhoudm@mail.ccnu.edu.cn}
\affiliation{Key Laboratory of Quark and Lepton Physics (MOE) and Institute of
            Particle Physics, Central China Normal University, Wuhan 430079,
            China}

\author{Hua Zheng}
\affiliation{School of Physics and Information Technology, Shaanxi Normal University, Xi'an 710119, China}

\author{Li-Lin Zhu}
\affiliation{College of Physics, Sichuan University, Chengdu 610064, China}

\author{Qiang Wang}
\affiliation{Key Laboratory of Quark and Lepton Physics (MOE) and Institute of
Particle Physics, Central China Normal University, Wuhan 430079,
            China}

\author{Yu-Liang Yan}
\affiliation{China Institute of Atomic Energy, P. O. Box 275 (10), Beijing 102413, China}

\author{Zhong-Qi Wang}
\affiliation{Nanjing University of Aeronautics and Astronautics, Nanjing 210000, China}

\author{Ben-Hao Sa}
\email[]{sabhliuym35@qq.com}
\affiliation{Key Laboratory of Quark and Lepton Physics (MOE) and Institute of Particle Physics, Central China Normal University, Wuhan 430079, China}
\affiliation{China Institute of Atomic Energy, P. O. Box 275 (10), Beijing 102413, China}

\date{\today}

\begin{abstract}
Two particle momentum correlation functions provide a nontrivial tool for
probing the strong interaction and/or extracting particle emission source
information in relativistic nuclear collisions. Although transport models can
describe the microscopic phase-space evolution of the collision system,
calculating correlation functions within the framework of transport models remains challenging.
In this paper, we employ the mixed-event technique to calculate two particle
momentum correlation function as $C(k^*)=
\mathcal{N} \xi(k^*)\frac{N_{\mathrm{same}}(k^*)}{N_{\mathrm{mixed}}(k^*)}$
based on the parton and hadron cascade model PACIAE simulated final hadronic
state (FHS) with introducing a modification factor $\xi(k^*)$ to
improve the treatment of final-state interactions and quantum statistics
effects in the PACIAE model. The simulated results show good agreement with
the ALICE data for $Kp$, $pp$, $p\Lambda$, and $\Lambda\Lambda$ momentum
correlation functions in $pp$ collisions at $\sqrt{s}=7$ TeV. On the other
hand, the particle emission source radius of the correlated pairs are
also evaluated based on the simulated FHS self-consistently. Since the PACIAE model
employs hadron-hadron cross sections derived from the additive quark model,
the calculation of two-particle momentum correlation functions
does not require prior assumptions about the interaction between the two correlated particles.
This successful ``PACIAE + modification factor" approach may shed light
on the future study of momentum correlation functions for dimesons, dibaryons, and even diexotic hadrons.
\end{abstract}

\maketitle

\section{Introduction}
The two particle momentum correlation
function (known as femtoscopy)~\cite{revw0,revw1,revw2,revw3,physrep99}
is defined as the ratio of the two particle momentum joint probability
to the product of corresponding single particle probability:
\begin{equation}
C(\boldsymbol{p}_1, \boldsymbol{p}_2) = \frac{P(\boldsymbol{p}_1, \boldsymbol{p}_2)}{P(\boldsymbol{p}_1) \cdot P(\boldsymbol{p}_2)}.
\label{funm1}
\end{equation}
Conventionally it was expressed approximately as the Koonin-Pratt formula~\cite{koonin,pratt}
\begin{equation}
C(\mathbf{k}^*)_{\text{KP}} = \int S(\mathbf{r}^*)|\Psi(\mathbf{r}^*, \mathbf{k}^*)|^2 \, d^3\mathbf{r}^*,
\label{funm3}
\end{equation}
in order to extract the particle emission source radius and the two
particle interaction parameters. In Eq.~(\ref{funm3}) the
$S(\mathbf{r}^*)$ is the assumed source distribution as function of
two particle relative distance $\mathbf{r}^*  = \mathbf{r}_1^* -
\mathbf{r}_2^*$ in the pair rest frame (PRF), and the
$\Psi(\mathbf{r}^*, \mathbf{k}^*)$ is two particle relative motion wave
function as function of the relative distance and the relative momentum
$ k^{*} = |\mathbf{p_1} - \mathbf{p_2}|/2 $ in the PRF.

Experimentally, the two particle momentum correlation function is measured as
the ratio of the number of samples of two particles with relative momentum of
$k^{*}$ counted in a single event ($N_{\text{same}}(k^{*})$) to the one
counted between two different events
($N_{\text{mixed}}(k^{*})$)~\cite{annrn21}, i.e.,
\begin{equation}
C(k^{*})_{\text{exp}} = \mathcal{N} \frac{{N_{\text{same}}}(k^{*})}
	                {{N_{\text{mixed}}}(k^{*})},
\label{funm2}
\end{equation}
where $\mathcal{N}$ is a normalization constant evaluated at the higher
relative momentum region where the correlation function is supposed to be unity.
In recent years, high-precision measurements at the RHIC and LHC have
expanded the scope of femtoscopy study to new particle pairs, such as
$\bar{p}\bar{p}$~\cite{ppbstar}, $p\Omega$, $p\Xi$~\cite{poalice},
and $\Lambda\Lambda$~\cite{llstar}. The three particle momentum
correlation function~\cite{pppalice} and the light nuclei momentum
correlation~\cite{ddalice} are also studied.

Theoretically, the various
frameworks~\cite{LLmod0,yka22,xeft,chir,julich,cats,amptfe,nnfe,kisi} ranging
from first-principle Lattice QCD to chiral effective field theories as well
as phenomenological models, have been applied to investigate the particle
momentum correlation functions. The Lednick$\mathrm{\acute{y}}$-Lyuboshitz (LL)
model~\cite{LLmod0,LLmod1} has emerged as a popular analytical framework for
the interpretation of experimental data, the parameterization final-state
interactions (FSI), and the extraction of the scattering parameters. While the
transport models can describe the microscopic phase space evolution of the
collision system, the study of momentum correlation functions in transport
model remains challenging due to the incomplete treatment of FSI and quantum
statistics (QS)~\cite{amptfe,iqmfe}.

In this work the mixed-event technique is proposed for the first time to
calculate two particle momentum correlation function as
\begin{equation}
C(k^*)=\mathcal{N}
\xi(k^*)\frac{N_{\mathrm{same}}(k^*)}{N_{\mathrm{mixed}}(k^*)},
\label{sa1}
\end{equation}
based on the PACIAE transport model simulated final hadronic state (FHS).
A modification factor $\xi(k^*)$ is introduced to improve the treatment of
the FSI and QS in the PACIAE model. This novel proposal is referred to as the
``PACIAE + modification factor" approach later. Except for the modification factor
$\xi(k^*)$, other variables in Eq.~(\ref{sa1}) are the same as the
corresponding ones in Eq.~(\ref{funm2}) with simulated FHS events instead
of the experimental events only.

\section{PACIAE model} The parton and hadron cascade model
PACIAE~\cite{paciae2,paciae3,paciae4} based on the
PYTHIA~\cite{pythia6,pythia8} is a Monte Carlo event generator
designed to simulate high-energy elementary particle and nucleus-nucleus collisions.
The collision process is composed of four
subprocesses: the partonic initialization, the parton cascade (partonic
rescattering), the hadronization, and the hadron cascade (hadronic
rescattering). In the first subprocess each proton-proton collision is
executed by PYTHIA model with presetting the string fragmentation turn-off
temporarily. Then the parton cascade is processed, where the Lowest-Order (LO)
perturbative quantum chromodynamics (pQCD) cross sections~\cite{combridge} is
employed. In the next step of hadronization, two models are available: the
Lund string fragmentation regime~\cite{pythia6} and the Monte Carlo coalescence
model~\cite{paciae2}. The former is adopted in this work. Finally, the
resulting hadrons undergo both hadronic rescattering (elastic or inelastic
2 $\to$ 2 process) and hadronic decay until the freeze-out of the final
hadronic state. We refer to the Ref.~\cite{paciae4} for the details.

In the context of hadronic rescattering, the total cross section of hadron $I$
bombarding with hadron $J$ is assumed to be proportional to the experimental
$NN$ total cross section~\cite{cross1,cross2} with a coefficient $C_{IJ}$
calculated using the additive quark model (AQM)~\cite{aqm,pythia8}:
\begin{equation}
C_{IJ}=\frac{n_{\text{eff}}^In_{\text{eff}}^J}{n_{\text{eff}}^Nn_{\text{eff}}^N},
\label{coef}
\end{equation}
\begin{equation}
n_{\text{eff}}^I=n_d^I+n_u^I+0.6n_s^I+0.2n_c^I+0.07n_b^I,
\label{coef2}
\end{equation}
where $n_i^I$ refers to the number of $i$-th valence quark (antiquark) in the
$I$-th hadron.

The PACIAE model has been successfully employed to reproduce experimental data
of the particle yield, the particle transverse momentum distribution, the rapidity
density distribution, and the exotic hadron production in the relativistic elementary and
heavy-ion collisions~\cite{xiez,leiw,caoj0,sexa,wangq,caoj1}.

\section{Modification factor} \label{th1}
Inspired by Refs.~\cite{LLmod0,LLmod1}, the modification factor in
Eq.~(\ref{sa1}) is empirically parameterized as
\begin{equation}
\xi(k^{*})
=
1+\lambda e ^ {-(Bk^{*})^{2}}
\left[
\delta_{C}\left(G(\eta)-1\right)
+
\frac{A}{1+(Bk^{*})^{2}}
\right],
\label{sa2}
\end{equation}
where $\lambda$ controls overall modification factor strength
and the parameter $\delta_{C}=+1,-1,0$ corresponds to repulsive,
attractive, absent Coulomb interaction, respectively.
The Gamow factor $G(\eta)$ reads
\begin{equation}
G(\eta)=
\frac{2\pi\eta}{e^{2\pi\eta}-1},
\qquad
\eta=
\frac{\alpha |Z_{1}Z_{2}|\mu}{k^{*}},
\end{equation}
where $\alpha$ is the fine-structure constant, $Z_1$ and $Z_2$ are the charge
numbers of the two correlated particles in units of electron charge of $e$,
and the $\mu= m_{1}m_{2}/(m_{1}+m_{2})$ is the reduced mass.
The rest parameters of $\lambda$, $A$, and $B$ in Eq.~(\ref{sa2}) are fixed by
fitting the $Kp$, $pp$, $p\Lambda$, and $\Lambda\Lambda$ momentum correlation
function to the corresponding ALICE data, respectively.

To effectively incorporate the FSI and QS effects beyond the transport model,
in the calculation of Eq.~(\ref{sa1}), each $N_{\mathrm{same}}(k^*)$ is
assigned a modification factor of $\xi(k^*)$.
We treat the Coulomb contribution exactly via the Gamow factor,
while the strong interaction is improved by constant $A$ modulated
by a Lorentzian factor $1/(1+(Bk^*)^2)$, which naturally vanishes at high $k^*$.

This ``PACIAE + modification  factor"
(``transport model + modification  factor") approach provides a
computationally efficient way to calculate two particle momentum correlation
function.

\begin{figure*}[htbp]
\includegraphics[scale=1.06]{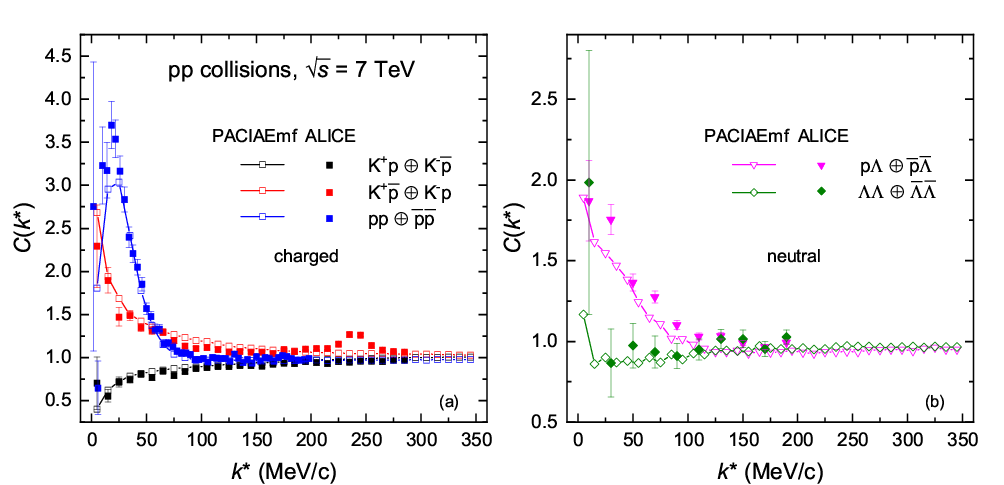}
\caption{The ``PACIAE + modification factor" approach simulated two particle
momentum corelation function (denoted as PACIAEmf) comparing with corresponding
ALICE data~\cite{kppb,plll} in $pp$ collisions at $\sqrt{s} = 7$ TeV: panel
(a) for charged $Kp$ and $pp$ pairs and (b) for $p\Lambda$ and
$\Lambda\Lambda$ neutral pairs.}
\label{ck5}
\end{figure*}

\begin{table*}[htpb]
\caption{The ``PACIAE + modification factor" approach simulated the effective radii of
the particle emission sources for $Kp$, $pp$,
$p\Lambda$, and $\Lambda\Lambda$ momentum correlation functions in $pp$
collisions at $\sqrt{s}=7$ TeV.}	
\centering
\renewcommand{\arraystretch}{1.5}
\begin{ruledtabular}
\begin{tabular}{cccccc}
Particle pairs &  $K^+p \oplus K^-\bar{p}$ & $K^+\bar{p} \oplus K^-p$ & $pp \oplus \bar{p}\bar{p}$ & $p\Lambda \oplus \bar{p}\bar{\Lambda}$ & $\Lambda\Lambda \oplus \bar{\Lambda}\bar{\Lambda}$ \\
$\langle R_{\text{evt}}\rangle$ (fm) &  $2.527\pm0.001$ & $2.517\pm0.001$ & $2.525\pm0.001$  & $3.211\pm0.013$  & $3.960\pm0.074$      \\

\end{tabular}
\end{ruledtabular}
\label{ta1}
\end{table*}

\section{Results and Conclusions}
The PACIAE 4.0 model~\cite{paciae4} is employed to generate 100 million events
for $pp$ collisions at $\sqrt {s}$ = 7 TeV with the same set of model
parameters as in Ref.~\cite{xiez}. Based on these events the two particle
momentum correlation function is calculated with the
``PACIAE + modification factor" approach above. The fitted $Kp$, $pp$,
$p\Lambda$, and $\Lambda\Lambda$ momentum correlation functions in $pp$
collisions at $\sqrt{s}=7$ TeV are shown in Fig.~\ref{ck5}. Here the kinematic
cuts in the transverse momentum ($p_{\rm T}$), the rapidity ($y$), and the
transverse sphericity ($S_{\rm T}$) are the same as the corresponding ones
in the ALICE experiments~\cite{kppb,plll} respectively.
In Fig.~\ref{ck5}, the ``PACIAE + modification factor" results generally reproduce the ALICE data for all particle pairs.
Strong enhancements at low $k^*$ are observed for the $K^+p$, $K^+\bar{p}$, $pp$, and $p\Lambda$ pairs,
while the $\Lambda\Lambda$ correlation remains close to unity.
The main features of the experimental data are well described.

On the other hand, an event-by-event effective radius estimator is
employed to calculate the radius of the correlated particle emission source
based on the particle spatial distributions in the simulated events above.
In a given event the $i-th$ correlated particle relative spatial distance
reads
\begin{equation}
r_i= \sqrt{(x_i-x_c)^2+(y_i-y_c)^2+(z_i-z_c)^2},
\end{equation}
where $(x_i, y_i, z_i)$ ($(x_c, y_c, z_c)$) refers to three coordinate of the
$i-th$ correlated particle (the spatial origin). To reduce the contribution
from low-probability events where correlated particles traveling too far away,
a weighting factor $w_i$ is introduced. The particle emission source radius in
this event is then defined as
\begin{equation}
R_\text{evt}=\frac{\sum_i w_i r_i}{\sum_i w_i}.
\label{sa3}
\end{equation}
The weighting factor is assumed as
\begin{equation}
w_i=
\max\left(0,\,1-\frac{r_i^2} {(r_\text{evt}^{\max})^2} \right),
\end{equation}
where $r_\text{evt}^\text{max}$ refers to the radius of the farthest
correlated particle in this event. By averaging over all events, the effective
radius of the particle emission sources is obtained and given in Table~\ref{ta1}.
The $Kp$ and $pp$ pairs exhibit similar source sizes,
while larger effective radii are obtained for the $p\Lambda$ and $\Lambda\Lambda$ pairs,
suggesting a more extended emission source for particle pairs involving $\Lambda$ hyperons.
Although the effective radius defined here is not identical to the Gaussian source radius used
in femtoscopic analyses, the corresponding source sizes for the $Kp$ and $pp$ pairs
are comparable to the experimentally adopted values~\cite{kppb,plll}.
Further experimental studies are desirable to test the source-size hierarchy predicted
by the present calculation.

In summary, the proposed ``PACIAE + modification factor" approach can
not only accurately calculate two particle momentum correlation function but
also provide the effective radius of particle emission source and without the
need to make prior assumption for the interaction between two correlated
particles. This approach would shed light on the future studies of the
dimeson, the dibaryon, and even the diexotic hadron momentum correlation
functions.

\section{Acknowledgments}
The authors thank Li-Sheng Geng, Xiao-Feng Luo, Xiao-Ming Zhang, Liang Zheng,
and Ke-Hao Zhang for helpful discussions.
This work of D.M.Z. is supported by the National Natural Science Foundation of
China under Grants No. 12375135, and by the 111 Project of the Foreign Expert
Bureau of China. W.C.Z. is supported by the Natural Science Basic Research
Plan in Shaanxi Province of China (Program No. 2023-JCYB-012). H.Z.
acknowledges the financial support from Key Laboratory of Quark and Lepton
Physics in Central China Normal University under grant No. QLPL2024P01.
Y.L.Y. acknowledges the financial support from the Key Laboratory of Quark and Lepton
Physics of the Central China Normal University under Grant No. QLPL201805 and
the Continuous Basic Scientific Research Project (Grant No. WDJC-2019-13).

\end{document}